\documentclass{aastex62}

\graphicspath{{./}{figures/}}

\shorttitle{Reverse Shock Emission from Short GRBs}
\shortauthors{Lloyd-Ronning}

\begin{document}

\title{Estimates of Reverse Shock Emission from Short Gamma-ray Bursts}

\correspondingauthor{Nicole Lloyd-Ronning}
\email{lloyd-ronning@lanl.gov}

\author{Nicole M. Lloyd-Ronning}
\affil{Center for Theoretical Astrophysics \& CCS-2, Los Alamos National Laboratory, Los Alamos, NM 87544}
\affil{University of New Mexico, Los Alamos, Los Alamos, NM 87544}



\begin{abstract}

 We investigate the expected radio emission from the reverse shock of short GRBs, using the afterglow parameters derived from the observed short GRB light curves. In light of recent results suggesting that in some cases the radio afterglow is due to emission from the reverse shock, we examine the extent to which this component is detectable for short GRBs. In some GRBs, the standard synchrotron shock model predicts detectable radio emission from the reverse shock when none was seen.  Because many physical parameters play a role in these estimates, our results highlight the need to more deeply explore the fundamental processes involved in GRB particle acceleration and emission.  However, with more rapid follow-up, we can test our standard model of GRBs, which predicts an early, radio bright reverse shock in many cases.
 
\end{abstract}



\section{Introduction} \label{sec:intro}

Perhaps the most robust model for short gamma-ray bursts (SGRBs) is the merger of two compact objects, such as two neutrons stars (NS-NS) or a neutron star and a black hole (NS-BH).  The timescales and energetics involved in the merger have always made this a plausible model for SGRBs \citep{Eich89, Nar92}, but other clues including the location of these bursts in their host galaxies, the lack of associated supernovae, and of course the recent detection of gravitational waves from a neutron star merger coincident with a SGRB \citep{Ab17} have provided convincing evidence that these bursts are associated with the older stellar populations expected of compact objects \citep{Fox05, Sod06a, Berg09, Koc10, LB10, Fong10, Berg10, Fong13, Fong14}. 

   There has been a concerted effort to follow up short GRBs with the goal of detecting the afterglow and potentially learning more about this class of gamma-ray bursts (for a review, see \cite{Berg14}).  To date, about $93\%, 84\%,$, and $58\%$ of SGRBs have been followed up in the X-ray, optical, and radio respectively \citep{Fong2015}. Of these follow-up efforts, $74\%$ have an X-ray afterglow, $34 \%$ have been seen in the optical, and only $7\%$ detected in the radio.
   
   Recently, \citet{LR17} investigated a sample of long GRBs that were followed up in the radio, and found bright bursts (with isotropic equivalent energy $E_{iso} > 10^{52}erg$) {\em without} radio afterglows had a significantly shorter intrinsic prompt duration.  They explored various reasons for the lack of afterglow in the context of different progenitor models; one possibility for the lack of radio afterglow is that this emission comes primarily from the reverse shock and that those with no a radio afterglow are in a parameter space with a weak reverse shock signal.  
   
  On the other hand, \citet{Las13, Las16} and \citet{Alex17} have recently reported the detection of a distinct reverse shock component in the afterglows of GRB130427A, GRB160509A, and 160625B.  They suggested that the external medium density must be low ($n < 1cm^{-3}$) in order to give a long-lived radio afterglow from the reverse shock (the low density allows for the emitting electrons to be in the so-called slow-cooling regime thereby giving rise to  longer-lived reverse shock emission). These results combined with those from \citet{LR17} prompted us to investigate why more short GRBs (with their presumed low circumstellar densities) do not have a detected radio afterglow from the reverse shock.  
   
     Using the multi-band afterglow fits from \citet{Fong2015}, we explore the detectability of the reverse shock component from SGRBs.  Using their fitted parameters for emission from the forward shock, we estimate the emission from the reverse shock, using the same formalism as \citet{Las13, Las16}. We find that in some cases (depending on the microphysical parameters), there should be a detectable radio signal at the time of the afterglow follow-up, when none was seen.  
     
     Our paper is organized as follows.  In Section 2 we describe how we calculate the radio flux from the forward and reverse shock using the standard formalism of synchrotron emission from a relativistic jet, using the fitted parameters from \cite{Fong2015}.  In Section 3 we present our results. We find that most of the reverse shock emission occurs too early to be detected in the radio, but in some cases this emission {\em should have been detected}. In Section 4, we summarize and present our conclusions.
 
 \section{Materials and Methods}

\citet{Fong2015} carried out an extensive effort, compiling all of the available afterglow data for 103 SGRBs, and fitting these data to the standard synchrotron forward external shock model.  Table 3 of \citet{Fong2015} gives the results of these fits - in particular, the values of $p$, $\epsilon_{B}$, the average isotropic kinetic energy $E_{iso}$, and the external density $n$ (assumed a constant, as expected for NS-NS or NS-BH progenitors).  Note that they assume the fraction of energy in the electrons is a fixed value of $\epsilon_{e}=0.1$. They performed two sets of fits to each burst, one in which the fraction of energy in the magnetic field $\epsilon_{B}$ is $0.1$ and one in which the value of $\epsilon_{B} = 0.01$.  If neither gave an acceptable fit, they allowed $\epsilon_{B}$ to be a free parameter (hence explaining the couple of entries with $\epsilon_{B} \neq 0.1$ or $0.01$). 

  We point out that four individual bursts were detected by \citet{Fong2015} in the radio band.  These bursts are GRB050724A, GRB051221A, GRB130603B, and GRB140903A. Table~\ref{tab:radiodetect} of this paper gives the time of observation in days and the flux in $\mu Jy$ detected at these times for these SGRBs. 

\begin{table}
	\centering
	\caption{Radio afterglow detections of short GRBs.}
	\label{tab:radiodetect}
	\begin{tabular}{lcr} 
		\hline
		GRB & $t_{obs}$ (days) & Flux ($\mu Jy$) \\
		\hline
		150724A & 0.57, 1.68   & 173, 465\\
		051221A & 0.91  & 155\\
		130603B & 0.37,1.43  & 125, 65\\
        140903A & 0.4,2.4,9.2 & 110, 187, 81 \\
       \hline
	\end{tabular}
\end{table}

\begin{figure*}
 \centering
	\includegraphics[width=6.0in]{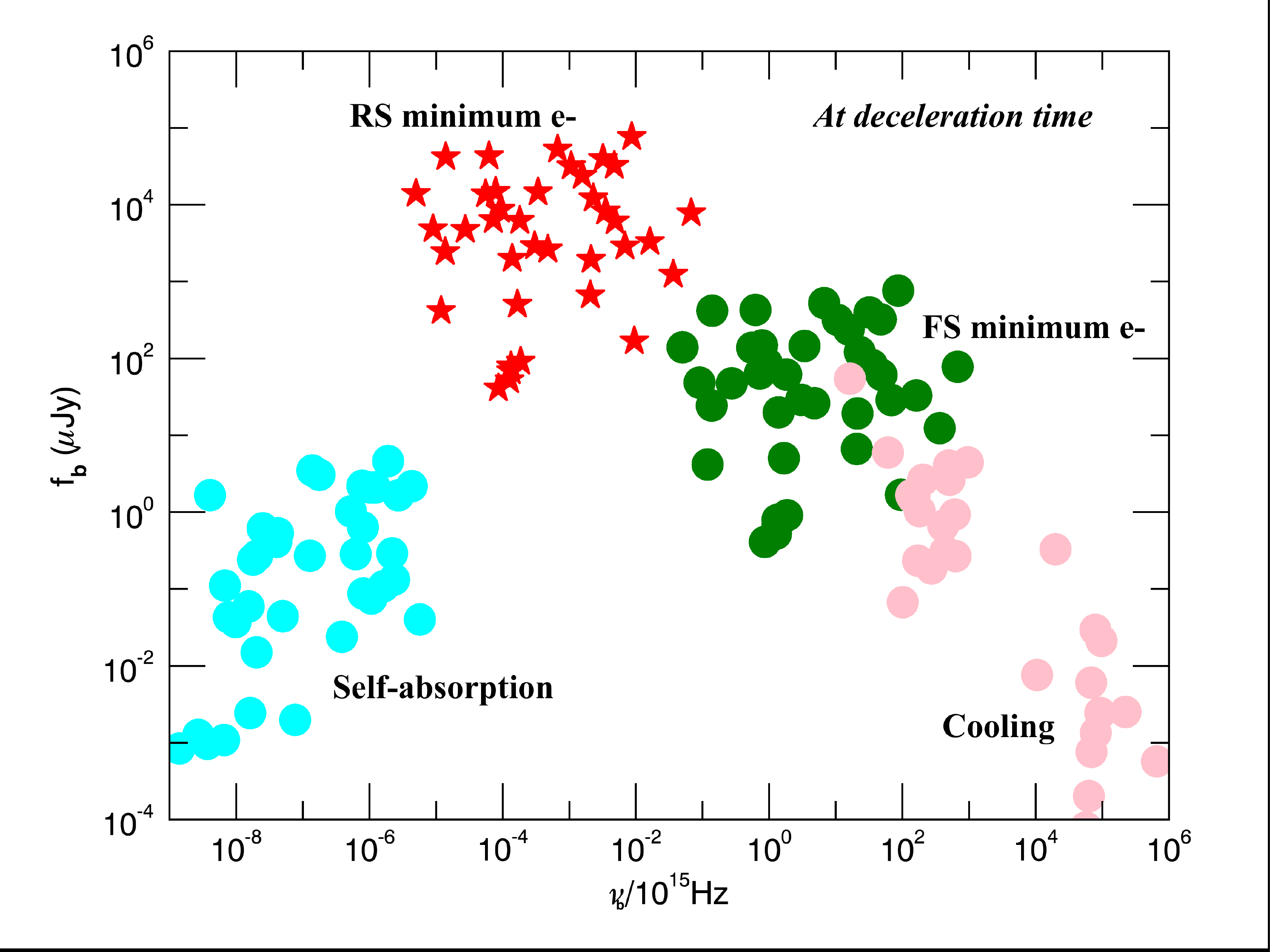}
    \caption{Flux at the characteristic frequency vs. characteristic frequency $\nu_{b}$ (normalized to $10^{15} Hz$ from a synchrotron spectrum in a standard external shock model, using data from \citet{Fong2015}, in which $\epsilon_{B} =0.01$ was employed in their fits.  The light blue dots indicate the forward shock self-absorption frequency $\nu_{a}$, the green dots show the frequency corresponding to the minimum or characteristic electron energy $\nu_{m}$, and the pink dots show the so-called cooling frequency $\nu_{c}$. The red stars indicate the minimum or characteristic electron frequency for the {\em reverse shock}, $\nu_{m,RS}$.} 
    \label{fig:freq}
\end{figure*}

In the standard picture of a relativistic external blast wave, the onset of the afterglow occurs around the deceleration time $t_{dec}$ - i.e. when the blast wave has swept up enough external material to begin to decelerate $t_{dec} \propto (E/n)^{1/3}\Gamma^{-8/3}$ \citep{BM76}, where $E$ is the energy in the blast wave, $n$ is the external particle number density and $\Gamma$ is the Lorentz factor of the blast wave.  One can calculate the characteristic synchrotron break frequencies at this time, depending on the global and microphysical parameters of the burst.   These expressions are given in Table 2 of \cite{GS02} for both a constant density and wind medium.  Figure ~\ref{fig:freq} shows the characteristic break frequencies (and the corresponding flux at these frequencies), using the parameters fitted from the \citet{Fong2015} data at the deceleration time (when the afterglow begins). The light blue dots indicate the self-absorption frequency $\nu_{a}$ of the forward shock, the green dots show the frequency corresponding to the minimum or characteristic electron energy $\nu_{m}$ of the forward shock, and the pink dots show the so-called cooling frequency $\nu_{c}$ of the forward shock (see, e.g. \citet{SPN98}, for more detailed explanations of these frequencies). In general, $\nu_{a} < \nu_{m} < \nu_{c}$ for the forward shock component.  The red stars indicate the minimum electron frequency for the {\em reverse shock}, $\nu_{m,RS} \approx \nu_{m}/\Gamma^{2}$ (note that this assumes the fraction of energy in the magnetic field is roughly the same for the forward and reverse shock, as explained below).  Again, to calculate both the characteristic frequencies and the fluxes at these frequencies, we employed the expression given in Table 2 of \cite{GS02}.      

\subsection{Jet Reverse Shock}
\label{sec:rs}

  There have been many studies of the reverse shock from a relativistic blast wave (e.g., \citet{MR97, SP99, Kob00, ZKM03, KZ03, ZWD05} and references therein), and the early-time radio flare observation of GRB 990123 has been attributed to the reverse shock \citep{KFS99, NP05}. In addition, \citet{SRR03} examined the expected strength of the reverse shock in six long GRBs, and were able to constrain the hydrodynamic evolution and bulk Lorentz factors of these bursts from this component. 
  
  As pointed out by these references and others, the evolution of the flux and break frequencies in the reverse shock depends on whether the blast wave is Newtonian or relativistic (among other factors), which in turn is related to the shell thickness $\Delta$ estimated from the observed duration $T$ by $\Delta \sim cT/(1+z)$. For a thick shell, $\Delta > l/2\Gamma^{8/3}$, where $l$ is the Sedov length in an interstellar medium $ \equiv (3E/4\pi n m_{p}c^{2})^{1/3}$, the reverse shock has time to become relativistic and the standard Blandford-McKee solution applies.  For a thin shell, the reverse shock remains Newtonian and the Lorentz factor of this shock evolves as $\Gamma_{RS} \sim r^{-g}$, with $g \sim 2$ \citep{Kob00}.  Short bursts with $T< 1s$ are likely in the thin shell - and therefore Newtonian - regime.  However, we note that for a range of $g$ values, the time evolution of the flux and characteristic frequencies are fairly similar between the relativistic and Newtonian regimes.  
  
  This standard treatment overly simplifies reverse shock emission by separating it into two distinct regimes (thick shell and thin shell), when in reality the shell thickness covers a range of values and could fall in between these regimes \citep{Kop15}. This simplified treatment also assumes that $\epsilon_{B}$ and $\epsilon_{e}$ are constant in the shell, which is not necessarily justified.  An evolving $\epsilon_{B}$ and $\epsilon_{e}$ will complicate the evolution of the flux and characteristic frequencies and allow additional degree of freedom in the treatment of the reverse shock.   
   
 However, generally speaking, because of the higher mass density in the shell, the peak flux in the reverse shock $f_{p, RS}$ will be higher by a factor of $\Gamma$ relative to the forward shock, 
  \begin{equation}
  f_{p, RS} \approx \Gamma f_{p, FS}
  \end{equation}
  but the minimum electron frequency in the reverse shock $\nu_{m, RS}$ will be lower by a factor of $\Gamma^{2}$,
  \begin{equation} 
  \nu_{m, RS} \approx \nu_{m, FS}/\Gamma^{2}
  \end{equation}
  assuming the forward and reverse shock have the same fraction of energy in the magnetic field (also not necessarily a well-justified assumption; see discussion below). For the purposes of comparing with others' analyses of reverse shock emission \citep{Las13, Las16}, we employ this prescription in our estimates below.       
  
 \subsubsection{Self-absorbed Reverse Shock}
 
  Because we are examining the radio emission, we need to be concerned with synchrotron self-absorption - under certain conditions lower energy photons are self-absorbed, and the flux is suppressed.  Self-absorption may be particularly relevant in the region of reverse shock, where the density is higher relative to the forward shock region. \citet{RZ16} calculated the relevance of the self-absorption frequency and flux in the reverse shock, before and after shock crossing. For our purposes - because we are looking at later time radio emission - we consider the frequencies and fluxes after the shock crosses the thin shell (but see their Appendix A.1 for expressions in all ranges of parameter space).  

\noindent Roughly, at the high radio frequencies we are considering here, the flux at the time of the peak can be obtained from equation 30 of \citet{RZ16}:
\begin{equation}
f_{p,RS} = f_{p, RS, \nu_{m}}(\nu_{a, RS}/\nu_{m, RS})^{-\beta}
\end{equation}
\noindent where $\beta=(p-1)/2$.\\
The reverse shock flux is suppressed at a minimum by factors ranging from about 0.3 to 0.01. We emphasize again, therefore, that our estimates are upper limits to the emission from the reverse shock.

\section{Results}

\begin{figure*} 
\centering
\includegraphics[width=3.2in]{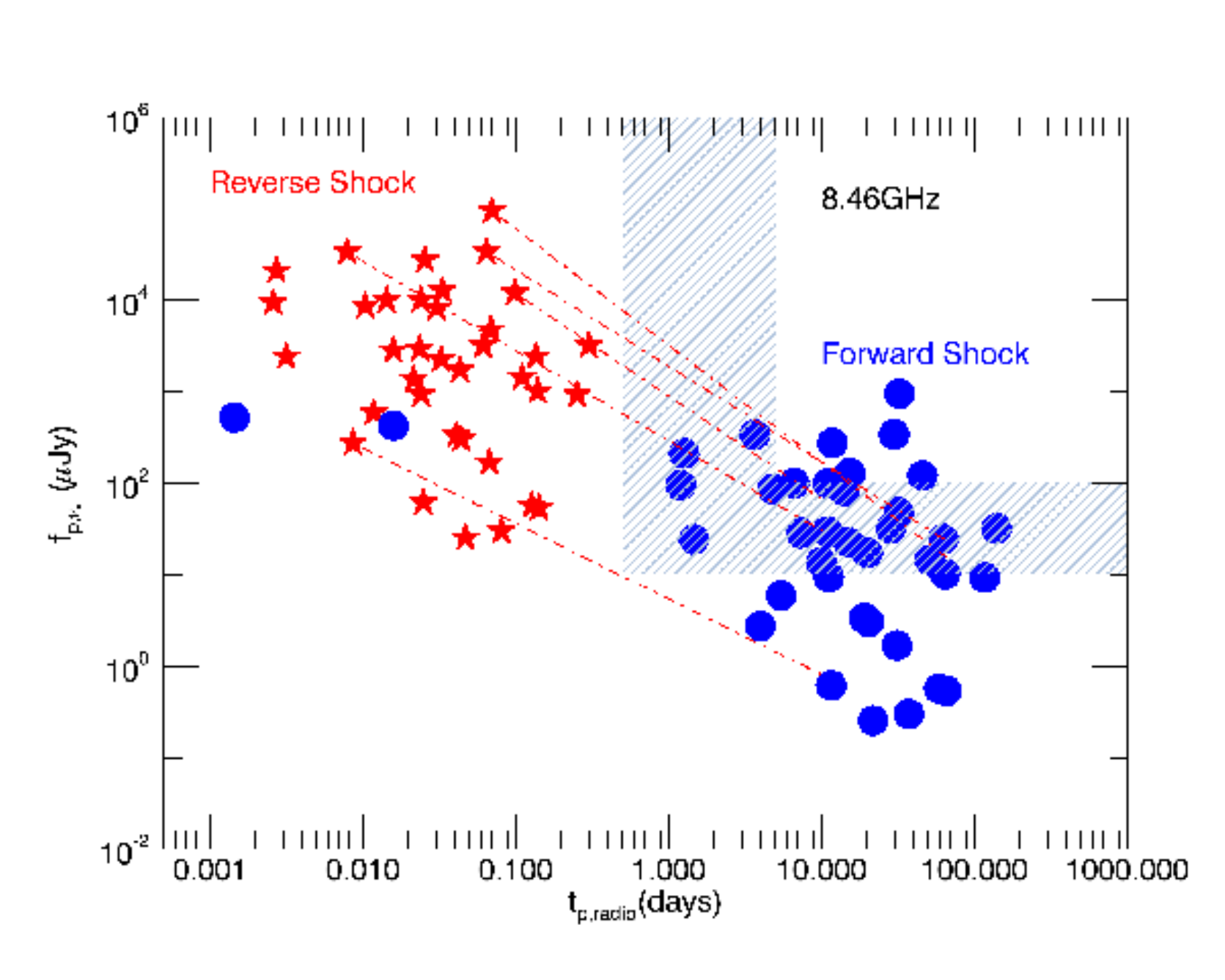}\includegraphics[width=3.2in]{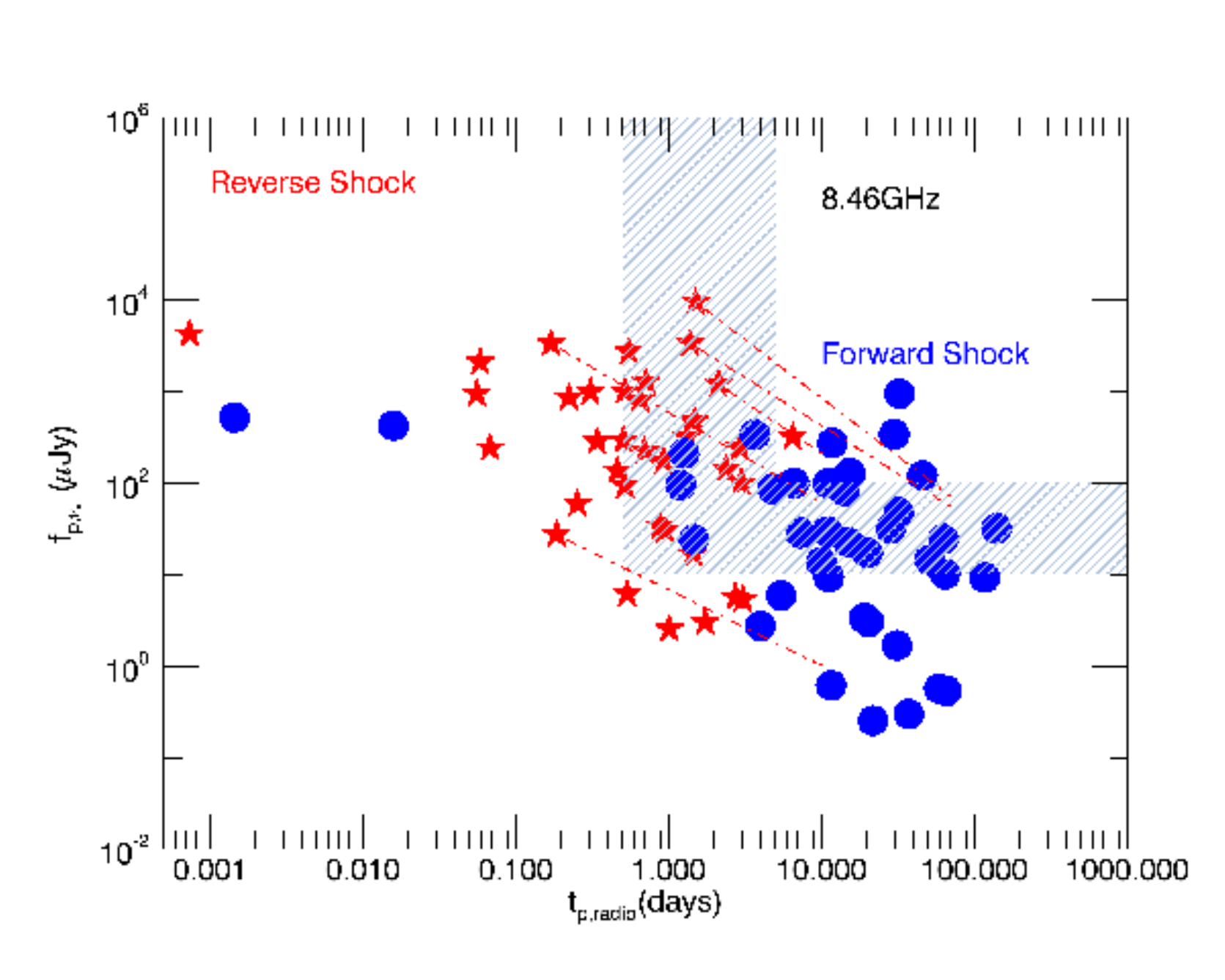}\\
\includegraphics[width=3.2in]{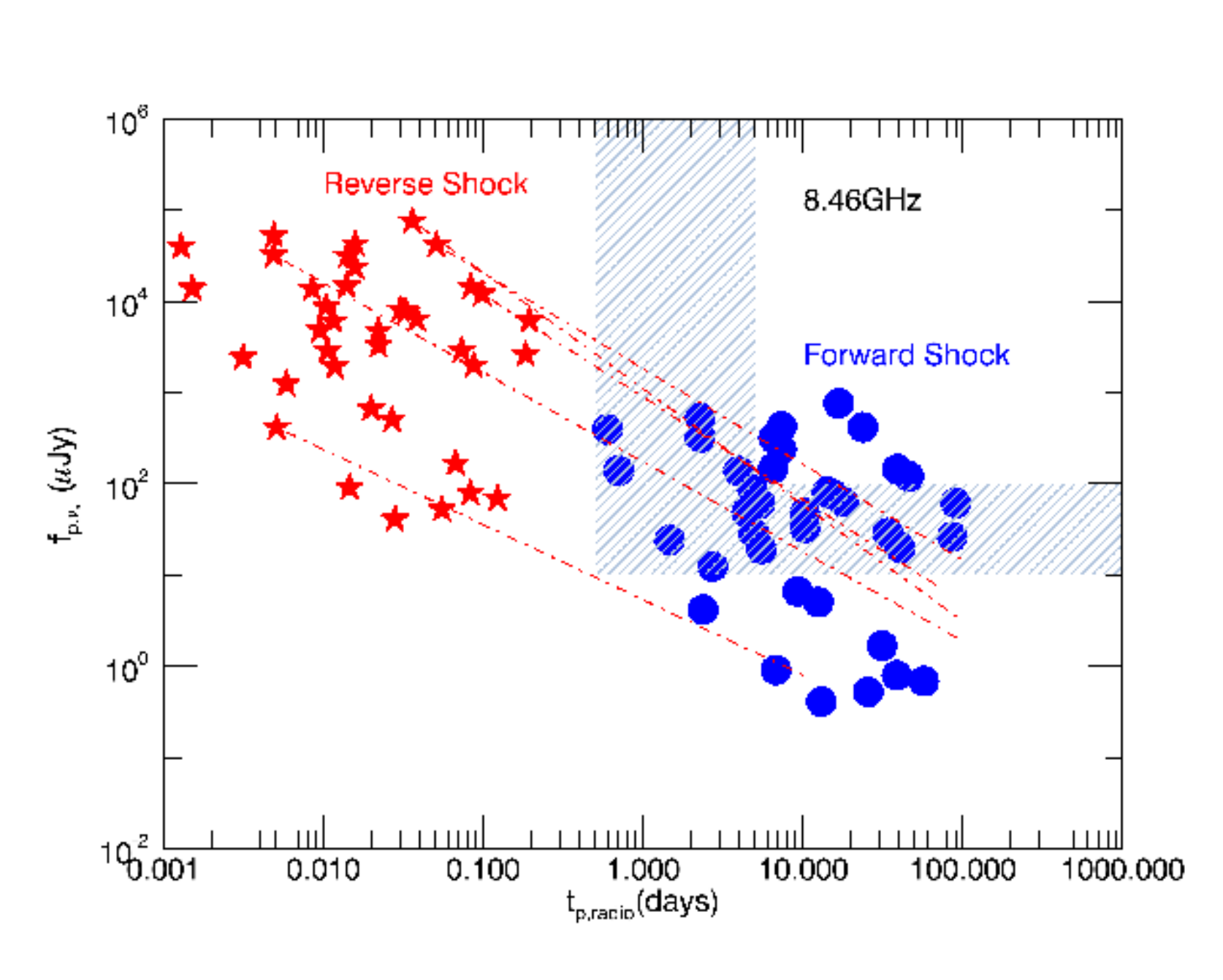}\includegraphics[width=3.2in]{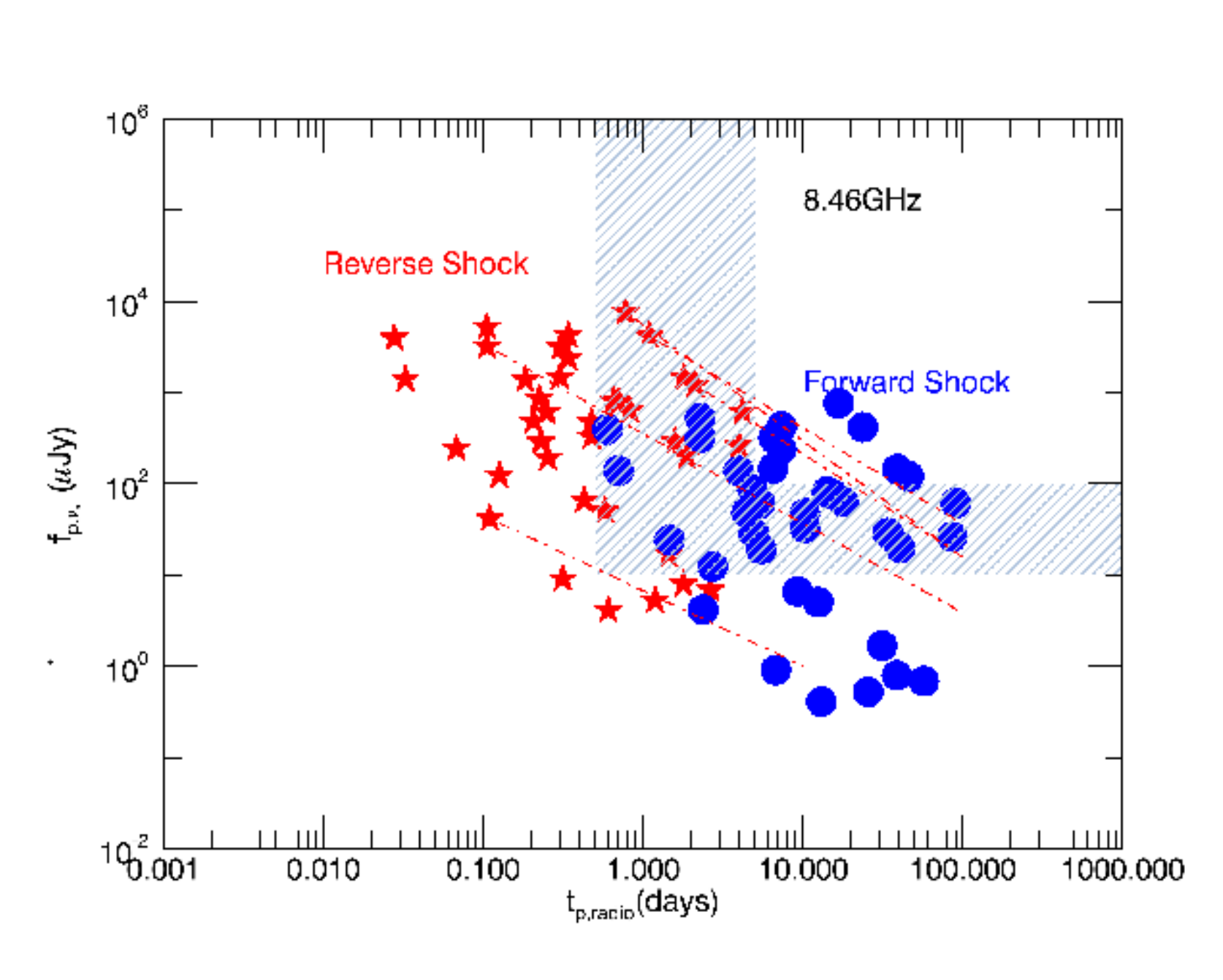}
\caption{Estimates of the peak flux from the forward (blue dots) and reverse (red stars) shock from synchrotron emission.  The vertical shaded region marks the temporal window when radio follow-up observations began for this sample.  The horizontal shaded region marks the rough observational flux density limit. {\em {\bf Top Left panel:}} $\epsilon_{B} = 0.1, \epsilon_{e} = \epsilon_{B}, \Gamma=100$.  {\em {\bf Top Right panel:}} $\epsilon_{B} = 0.1, \epsilon_{e} = \epsilon_{B}, \Gamma=10$. The red dashed lines show the flux decay as a function of time for a few representative bursts. {\em {\bf Bottom Left panel:}} $\epsilon_{B} = 0.01, \epsilon_{e} = 0.1, \Gamma=100$.  {\em {\bf Bottom Right panel:}} $\epsilon_{B} = 0.01, \epsilon_{e} = 0.1, \Gamma=10$. The red dashed lines show the predicted flux decay as a function of time for a few representative bursts.}
\label{fig:fpFSRSa}
\end{figure*}

 Figure~\ref{fig:fpFSRSa} shows our estimates of the peak flux at $\nu_{m}$ for the forward (blue circles) and reverse (red stars) shocks at the time $\nu_{m}$ reaches the radio band of 8.46GHz. The left panels are for a Lorentz factor $\Gamma=100$, while the right panels are for $\Gamma=10$.  The top panels of Figure~\ref{fig:fpFSRSa} utilize the $\epsilon_{B} = 0.1$ fits from \citet{Fong2015}, while the bottom panels utilize the parameters from their $\epsilon_{B} = 0.01$ fits.  Note that \citet{Fong2015} report the median of the observing time response for the radio afterglow follow-up observations to be about $1$ day.  This is reflected in Figure~\ref{fig:fpFSRSa} by the vertical shaded regions.  The horizontal shaded regions show roughly the detector flux limits.  The red dashed lines show the standard synchrotron flux decay as a function of time for a few representative bursts, assuming the reverse shock has become relativistic and a Blandford-McKee solution applies. This temporal decay is computed using the fitted parameters of \cite{Fong2015} (which determine the relative values of the characteristic synchrotron break frequencies) and the expressions for synchrotron flux given in Table 2 of \cite{GS02}.   
 
 We point out that although many sGRBs have apparent non-thermal gamma-ray photons that constrain the Lorentz factor to be large, $\Gamma \geq 100$  (a compact region will be optically thick to pair production at gamma-ray energies, unless the region is moving relativistically, \cite{LS01}), some sGRBs do not impose such stringent constraints, and a $\Gamma \sim 10$ is sufficient to allow for their non-thermal spectra (the most famous example is GRB170817 \cite{Ab17}, but see also \cite{Bur17} which show a sample of GRBs with a lack of high energy photons).  We display both $\Gamma = 100$ and $\Gamma = 10$ not necessarily to argue for low Lorentz factor sGRBs but to show how the reverse shock flux and its peak time vary as a function of Lorentz factor.
 
 It is clear that in the context of this model, many of the reverse shock bursts are missed because they tend to peak before the beginning of the observations.   

Note that a few bursts that went undetected in the radio (i.e. not one of the four listed in Table~\ref{tab:radiodetect}) indicate a forward shock component in the observational window in Figure~\ref{fig:fpFSRSa}. 
However, on closer examination, comparing the time of observations of these particular bursts to the predicted time of the peak (at $\nu_{m}$), we see that the radio observations occurred {\em well before} the predicted peak time (which occurs at $ \approx 10$'s of days in all of our models), and may be why it was not detected.  
However, as discussed above, {\em the reverse shock emission falls above the flux limit} for several bursts (in particular for GRB11112A, GRB121226A, GRB131004A, GRB150101B) during the time of their radio observations, particularly for the lower Lorentz factor cases ($\Gamma = 10$; right panels of Figure~\ref{fig:fpFSRSa}).  The fact that this emission was not detected suggests that - at least in some cases - the reverse shock flux derived from this standard prescription of GRB afterglow emission is overly simplistic and give misleading values for the flux (we again emphasize that we are looking in the optically thin limit here and it may also be that the reverse shock emission was self-absorbed in these cases).

  In any case, it is clear that rapid follow-up in the radio gives us a better chance to detect and/or constrain this important component, potentially breaking some of the degeneracies amongst the physical parameters in the models and allowing us to better understand the physics behind SGRB emission. 
\section{Conclusions}

 We have investigated radio emission from short gamma-ray bursts, using fits from existing broadband afterglow data \citep{Fong2015} in the context of the standard synchrotron shock model for GRB emission. In particular, we have looked at the peak flux from the forward and reverse shock components of the relativistic jet. We find in some cases that the reverse shock component should have been detected in the context of this standard model.  The lack of detection suggests any number of oversimplifications in the model, including potentially variable microphysical parameters, a mis-estimated bulk Lorentz factor, and/or not properly accounting for self-absorption.\\

We can get additional important information on short gamma-ray bursts if there is rapid follow-up ($< 1 day$) in the radio - this will give the best chance of detecting the reverse shock emission component. High Lorentz factor outflows $\Gamma \sim 100$ peak very early ($t \sim 0.05$ day) and may be very challenging to detect. Lower Lorentz factor outflows $\Gamma \sim 10$ peak later and give us a better chance of temporally catching the reverse shock; however, the flux will be lower for the less relativistic outflows.  The circumburst density must also be low enough to allow for a slow-cooling reverse shock (as mentioned in \citet{Las13, Las16}), but such densities are expected for compact object binary progenitors of sGRBs.\\

The electromagnetic signals can be very sensitive to the values of the microphysical parameters, such as the fraction of energy in the electrons and magnetic field, so a concerted effort to a concerted effort to more definitively constrain those parameters from a theoretical standpoint would be helpful in breaking the degeneracies and pinning down global burst parameters like kinetic energy, circumburst density, etc. These latter parameters can help constrain the progenitor.

Once again, more rapid follow-up (ideally within hours) with greater sensitivity could produce significant number of detections of radio emission from the reverse shock. A lack of detection would also constrain models to some extent and point us toward areas in which we are oversimplifying our treatment of GRB emission. 
 
 We point out again, however, that the radio emission is just one piece of the puzzle in understanding GRB emission and it is only through multi-wavelength follow-up that we will really be able to constrain the underlying physics of the outflow producing gamma-ray bursts.  Efforts in this vein are particularly timely in light of the near era of gravitational wave detection from a double neutron star merger.  A better understanding of the various components of electromagnetic emission from these objects will provide a more complete picture of these systems and ultimately help us understand their role in the context of stellar evolution in the universe. 

\vspace{6pt} 



\acknowledgments{Work at LANL was done under the auspices of the National Nuclear Security Administration of the US Department of Energy at Los Alamos National Laboratory LA-UR-17-24900.}



\end{document}